\documentclass[a4paper,12pt]{article}
\usepackage{amsfonts}
\usepackage{latexsym}
\usepackage{amssymb}
\usepackage{color}
\date{}

\begin{document}

\title{A modified Belinfante/Rosenfeld Procedure\\for Testing the Compatibility of
\\General-Covariant Continuum Physics and\\General Relativity Theory
}

\author{W. Muschik\footnote{Corresponding author:
muschik@physik.tu-berlin.de}
\\
Institut f\"ur Theoretische Physik\\
Technische Universit\"at Berlin\\
Hardenbergstr. 36\\D-10623 BERLIN,  Germany}
\maketitle

            \newcommand{\be}{\begin{equation}}
            \newcommand{\beg}[1]{\begin{equation}\label{#1}}
            \newcommand{\ee}{\end{equation}\normalsize}
            \newcommand{\bee}[1]{\begin{equation}\label{#1}}
            \newcommand{\bey}{\begin{eqnarray}}
            \newcommand{\byy}[1]{\begin{eqnarray}\label{#1}}
            \newcommand{\eey}{\end{eqnarray}\normalsize}
            \newcommand{\beo}{\begin{eqnarray}\normalsize}
            \newcommand{\R}[1]{(\ref{#1})}
            \newcommand{\C}[1]{\cite{#1}}

            \newcommand{\mvec}[1]{\mbox{\boldmath{$#1$}}}
            \newcommand{\x}{(\!\mvec{x}, t)}
            \newcommand{\m}{\mvec{m}}
            \newcommand{\F}{{\cal F}}
            \newcommand{\n}{\mvec{n}}
            \newcommand{\argm}{(\m ,\mvec{x}, t)}
            \newcommand{\argn}{(\n ,\mvec{x}, t)}
            \newcommand{\T}[1]{\widetilde{#1}}
            \newcommand{\U}[1]{\underline{#1}}
            \newcommand{\X}{\!\mvec{X} (\cdot)}
            \newcommand{\cd}{(\cdot)}
            \newcommand{\Q}{\mbox{\bf Q}}
            \newcommand{\p}{\partial_t}
            \newcommand{\z}{\!\mvec{z}}
            \newcommand{\bu}{\!\mvec{u}}
            \newcommand{\rr}{\!\mvec{r}}
            \newcommand{\w}{\!\mvec{w}}
            \newcommand{\g}{\!\mvec{g}}
            \newcommand{\D}{I\!\!D}
            \newcommand{\se}[1]{_{\mvec{;}#1}}
            \newcommand{\sek}[1]{_{\mvec{;}#1]}}            
            \newcommand{\seb}[1]{_{\mvec{;}#1)}}            
            \newcommand{\ko}[1]{_{\mvec{,}#1}}
            \newcommand{\ab}[1]{_{\mvec{|}#1}}
            \newcommand{\abb}[1]{_{\mvec{||}#1}}
            \newcommand{\td}{{^{\bullet}}}
            \newcommand{\eq}{{_{eq}}}
            \newcommand{\eqo}{{^{eq}}}
            \newcommand{\f}{\varphi}
            \newcommand{\dm}{\diamond\!}
            \newcommand{\seq}{\stackrel{_\bullet}{=}}
            \newcommand{\st}[2]{\stackrel{_#1}{#2}}
            \newcommand{\om}{\Omega}
            \newcommand{\emp}{\emptyset}
            \newcommand{\bt}{\bowtie}
            \newcommand{\btu}{\boxdot}
\newcommand{\Section}[1]{\section{\mbox{}\hspace{-.6cm}.\hspace{.4cm}#1}}
\newcommand{\Subsection}[1]{\subsection{\mbox{}\hspace{-.6cm}.\hspace{.4cm}
\em #1}}

\newcommand{\const}{\textit{const.}}
\newcommand{\vect}[1]{\underline{\ensuremath{#1}}}  
\newcommand{\abl}[2]{\ensuremath{\frac{\partial #1}{\partial #2}}}

\noindent
{\bf Abstract} 
Creating a modified Belinfante/Rosenfeld procedure, Mathisson-Papapetrou-like equations
are derived by which a comparison of General-Covariant Continuum Physics with
General Relativity Theory becomes possible.  
\vspace{.4cm}\newline

\section{Introduction}

The Belinfante/Rosenfeld procedure generates a symmetric and divergence-free tensor
of second order which serves as a source of Einstein's equations. The original procedure
implemented for the special-relativistic case \C{B6,B7} was recently extended to
General-Covariant Continuum Physics (GCCP) \C{A}. The special-relativistic 
Belinfante/Rosenfeld procedure and also the general-covariant one start out with
an especially defined combination of spin divergences which together with the
symmetric part of the energy-momentum tensor is symmetrized and made divergence-free. Special constraints by performing the procedure do not appear in the special-relativistic case, whereas in the general-covariant case such constraints are generated: the Mathisson-Papapetrou equations which have to be satisfied as necessary conditions
for the energy-momentum tensor and the spin tensor of GCCP \C{A}.
\vspace{.3cm}\newline
Now the question arises whether the general-covariant Belinfante/Rosenfeld  procedure
is unique, or if there exist different combinations of spin divergences for generating
sym\-me\-tric and divergence-free tensors. This question
is here investigated for the case of GCCP by introducing a family of Belinfante/Rosenfeld  procedures.
\vspace{.3cm}\newline
The paper is organized as follows: After this introduction and recalling the
general-covariant Belinfante/Rosenfeld procedure, a family of spin divergences is introduced, implementing a modified Belinfante/Rosenfeld  procedure resulting in
Mathisson-Papapetrou-like equations. These equations decompose into two classes:
one class contains the curvature tensor for special family parameters , the other does not.
The first class is chosen because we are interested in the effectiveness of the 
modified Belinfante/Rosenfeld procedure in General Relativity Theory
(GRT) with regard to GCCP. Inserting the
balance equations of energy-momentum and spin into the necessary
Mathisson-Papapetrou-like equations, we obtain a system of
differential equations for the spin and the metric.  Finally, a remark is made about the compatibility of GCCP with GRT with regard to constitutive properties.

\section{The Mathisson-Papapetrou equations\label{MPE}}

First of all, we start out with the general-covariant Belinfante/Rosenfeld procedure
citing the well known\newline
$\blacksquare$ {\sf Proposition}\C{A,C1,C2}: 
The general-covariant Belinfante/Rosenfeld procedure generates a symmetric and divergence-free
tensor 
\bee{U1}
^\dagger\Theta^{ab}\ :=\ T^{ab}
-\frac{1}{2}\Big[S^{cab}+S^{abc}+S^{bac}\Big]_{;c},
\ee
if  the Mathisson-Papapetrou equations
\bee{U2}  
\frac{1}{2}S^{cab}{_{;c}}\ =\ T^{[ab]},\qquad
T^{ab}{_{;a}}\ =\ \frac{1}{2}\Big[S^{cab}+S^{abc}+S^{bac}\Big]_{;c;a}\ =\
-\frac{1}{2}R^b_{klm}S^{klm}
\ee
are valid as necessary constraints.\hfill $\blacksquare$\newline
Here, $T^{ab}$  is the in general non-symmetric and not divergence-free 
energy-momentum tensor, $S^{cab}$ the spin tensor and $R^b_{klm}$
the curvature tensor. 
The Mathisson-Papapetrou equations \R{U2} are general-covariant including the
special-relativistic case which is characterized by replacing the covariant derivatives by
commuting partial ones and by $R^b_{klm}\equiv 0$.
Subtracting \R{U2}$_1$ from \R{U1} results in
\bee{U4}
^\dagger\Theta^{ab}\ =\ T^{(ab)} -\frac{1}{2}\Big[S^{abc}+S^{bac}\Big]_{;c}\ =\
T^{(ab)}-S^{(ab)c}{_{;c}},
\ee
a tensor which is symmetric and divergence-free according to \R{U1} and \R{U2}$_2$.
Consequently, the general-covariant Belinfante/Rosenfeld procedure
transforms by use of the
spin divergences a not necessary symmetric and divergence-free tensor into such one
\bee{aU4}
T^{ab} \neq T^{ba},\ T^{ab}{_{;a}} \neq 0 \quad\st{{BRP}}{\longrightarrow}\quad
^\dagger\Theta^{ab} = ^\dagger\Theta^{ba},\ ^\dagger\Theta^{ab}{_{;a}}=0,
\ee
starting out with the definition \R{U1}.
\vspace{.3cm}\newline
The question now arises, whether symmetrizations are possible which do not refer to
such ad-hoc setting \R{U1}, and by which conditions the general-covariant
Belinfante/Rosenfeld procedure \R{aU4} is characterized. Consequently, we are looking
for another possibility than
\R{U1} to use spin divergences for a symmetrization of the energy-momentum tensor.

\section{A modified Belinfante/Rosenfeld Procedure}
\subsection{A family of spin divergences}

The general-covariant Belinfante/Rosenfeld procedure starts out with the ad-hoc
expression \R{U1}. We now replace the square bracket in \R{U1} by an ad-hoc family of
spin divergences whose family parameters are scalars $(\lambda,\mu,\nu)$ 
implementing a modified Belinfante/Rosenfeld  procedure. Analogously to \R{U1},
we define
\byy{W1}
\Theta^{ab}(\lambda,\mu,\nu)&:=& T^{ab}-\Sigma^{cab}{_{;c}}(\lambda,\mu,\nu)
\\ \nonumber
\Sigma^{cab}(\lambda,\mu,\nu) &:=& \mu S^{cab} +\lambda S^{(ab)c}
+\nu S^{[ab]c}\ =\
\hspace{2.5cm}
\\ \nonumber
&=& \mu S^{cab}+\lambda\frac{1}{2}(S^{abc}+S^{bac})
+\nu\frac{1}{2}(S^{abc}-S^{bac})\ =\
\\ \label{aQ7}
&=& \mu S^{cab}+\frac{1}{2}(\lambda+\nu)S^{abc}
+\frac{1}{2}(\lambda-\nu)S^{bac}.
\vspace{.3cm}\eey
According to its definition \R{aQ7}$_1$, the symmetric and anti-symmetric parts of
$\Sigma^{cab}(\lambda,\mu,\nu)$ are\footnote{The
semicolon denotes covariant derivatives, round brackets
the symmetric part of a tensor, square brackets its anti-symmetric part.}
\byy{W2}
\Sigma^{c(ab)}(\lambda) &=& \lambda S^{(ab)c},
\\ \label{W2a}
\Sigma^{c[ab]}(\mu,\nu) &=& \mu S^{cab} +\nu S^{[ab]c},
\eey
by taking the anti-symmetry of the spin tensor
\bee{W3}
S^{cab}\ =\ -S^{cba}
\ee
into account. We obtain from \R{aQ7}$_3$ by changing $c\leftrightarrow a$ and
using \R{W3}
\bey\nonumber
\Sigma^{acb}(\lambda,\mu,\nu) &=&
\mu S^{acb}+\frac{1}{2}(\lambda+\nu)S^{cba}+\frac{1}{2}(\lambda-\nu)S^{bca}\
=
\\ \label{W4}
&=&
-\mu S^{abc}-\frac{1}{2}(\lambda+\nu)S^{cab}-\frac{1}{2}(\lambda-\nu)S^{bac}.
\eey
Addition of \R{W4}$_2$ with \R{aQ7}$_3$ results in
\bee{W5}
\Sigma^{cab}(\lambda,\mu,\nu)+\Sigma^{acb}(\lambda,\mu,\nu) \!\! =\!\!
\Big[\mu-\frac{1}{2}(\lambda+\nu)\Big]S^{cab} + \Big[\frac{1}{2}(\lambda+\nu)-\mu\Big]S^{abc}.
\ee
Consequently, we obtain a condition for the family parameters generating the
anti-symmetry of $\Sigma^{cab}(\lambda,\mu,\nu)$ in the first two indices
\bee{W7} 
\mbox{if}\quad 2\mu=\lambda+\nu\quad\longrightarrow\quad 
\Sigma^{cab}(\lambda,\mu,\nu)\ =\
-\Sigma^{acb}(\lambda,\mu,\nu).
\ee
According to \R{W5},
no anti-symmetry of $\Sigma^{cab}(\lambda,\mu,\nu)$ exists, if the family parameters
do not obey \R{W7}$_1$.

\subsection{Symmetrization procedure}

Starting out with \R{W1}, we demand\footnote{The sign $\st{\td}{=}$ stands for a
setting.}
\byy{W9}
\Theta^{[ab]}(\lambda,\mu,\nu)&=& T^{[ab]}
-\Sigma^{c[ab]}{_{;c}}(\lambda,\mu,\nu)\ \st{\td}{=}\ 0,
\\ \label{W10}
\Theta^{ab}{_{;a}}(\lambda,\mu,\nu)&=&
T^{ab}{_{;a}}-\Sigma^{cab}{_{;c;a}}(\lambda,\mu,\nu)\ \st{\td}{=}\ 0,
\eey
generating a symmetric and divergence-free tensor $\Theta^{ab}$.
Consequently, we obtain from \R{W9}$_2$ and \R{W2a}
\bee{W11}
T^{[ab]}\ =\ \Sigma^{c[ab]}{_{;c}}(\mu,\nu)\ =\ \mu S^{cab}{_{;c}}
+\nu S^{[ab]c}{_{;c}}.
\ee
and taking \R{W9}$_2$, \R{W10}$_2$ and \R{W2} into account, \R{W1} results in
\byy{W12}
\Theta^{ab}(\lambda) &=& T^{(ab)} - \Sigma^{c(ab)}{_{;c}}(\lambda)\ =\ T^{(ab)}
-\lambda S^{(ab)c}{_{;c}},
\\ \label{W13}
\Theta^{ab}{_{;a}}(\lambda) &=& 0\ =\ T^{(ab)}{_{;a}}-\lambda S^{(ab)c}{_{;c;a}}.
\eey

\subsection{Mathisson-Papapetrou-like equations and curvature}

A comparison of \R{U2}$_{1,2}$ with \R{W11} and \R{W10} turns out that \R{W10}
becomes according to \R{aQ7}$_3$ and \R{W11}
\bee{W14}
T^{ab}{_{;a}}\ =\ \Sigma^{cab}{_{;c;a}}(\lambda,\mu,\nu)\ =\
\Big[\mu S^{cab}+\frac{1}{2}(\lambda+\nu)S^{abc}
+\frac{1}{2}(\lambda-\nu)S^{bac}\Big]_{;c;a}.
\vspace{.3cm}\ee
According to \R{U2},
\R{W11} and \R{W14} are Mathisson-Papapetrou-like equations which change into the
original ones, if the family parameters are
\bee{W16}
\lambda\ =\ 1,\qquad \mu\ =\ 1/2,\qquad \nu\ =\ 0.
\ee
This combination of the family parameters satisfies \R{W7}$_1$ so that the
anti-symmetry
\bee{W17}
\Sigma^{cab}(1,1/2,0)\ =\ -\Sigma^{acb}(1,1/2,0)
\ee
is valid in this case.
\vspace{.3cm}\newline
We now remember the\newline
$\blacksquare$ {\sf Proposition}\C{A}:
If $\Sigma^{cab}$ is anti-symmetric in the first two indices
\bee{W18}
\Sigma^{cab}(2\mu-\nu,\mu,\nu)\ =\ -\Sigma^{acb}(2\mu-\nu,\mu,\nu)
\ee
according to \R{W7}, the second derivatives in \R{W14} can be replaced by the
curvature tensor $R^b_{klm}$
\bey\nonumber
T^{ab}{_{;a}} &=& \Sigma^{cab}{_{;c;a}}(2\mu-\nu,\mu,\nu)\ =\ 
\\ \label{W19}
&=& \Big[\mu S^{cab}+\mu S^{abc}
+(\mu-\nu)S^{bac}\Big]_{;c;a}\ =\ -R^b_{klm}\Sigma^{k[lm]}(\mu,\nu).
\eey
This replacement is only possible, if \R{W18} is valid, otherwise the second derivatives
in \R{W14} cannot be replaced by the curvature tensor.\hfill$\blacksquare$\newline
Inserting \R{W2a}, we obtain from \R{W19} and \R{W11} the
Mathisson-Papapetrou-like equations
\bee{W20}
T^{ab}{_{;a}}\ =\ -R^b_{klm}\Big(\mu S^{klm}+\nu S^{[lm]k}\Big),\qquad
T^{[ab]}\ =\ \mu S^{cab}{_{;c}}+\nu S^{[ab]c}{_{;c}}.
\ee
These Mathisson-Papapetrou-like equations represent the energy-momentum balance
equation \R{W20}$_1$ and the spin balance equation \R{W20}$_2$ which both are
necessary for generating the symmetric and divergence-free tensor \R{W12}, if using
\R{W18} in \R{W1}. The curvature tensor $R^b{_{klm}}$ is determined by the
space-time geometry which is choosed by a back-ground metric or by corresponding
field equations\footnote{Especially here, we will choose Einstein's equations in
sect.\ref{GRT}}.
The corresponding symmetric and divergence-free tensor \R{W12} is according to
\R{W7}$_1$
\bee{W22}
\Theta^{ab}(2\mu-\nu)\ =\ T^{(ab)}-(2\mu-\nu)S^{(ab)c}{_{;c}}.
\ee
Choosing the parameters \R{W16} results in the Mathisson-Papapetrou equations
\R{U2}. 
\vspace{.3cm}\newline
For generating the Mathisson-Papapetrou equations two facts are
necessary: One has to know the relation between the anti-symmetric part of the
energy-momentum tensor and the spin divergences \R{W20}$_2$\footnote{a
knowledge from
outside the Belinfante/Rosenfeld procedure which establish the factor 1/2 in \R{U1}},
and \R{W18} has to be valid, because otherwise the curvature tensor cannot be
introduced to the Mathisson-Papapetrou-like equation \R{W20}$_1$. Consequently,
the applicability of the Belinfante/Rosenfeld procedure depends on knowledge from the
outside of the procedure: the scalars $\mu$ and $\nu$ have to be implemented by
external facts, such as \R{W16} which are Lagrangian-based. 
\vspace{.3cm}\newline
We now consider the case for which the energy-momentum tensor itself is symmetric and
divergence-free
\bee{W22a}
T^{[ab]}\ =\ 0,\qquad T^{ab}{_{;a}}\ =\ 0,
\ee
the case for which the Belinfante/Rosenfeld procedure is dispensible because the
energy-momentum tensor itself satisfies the result of this procedure.
According to \R{W22a}$_1$, \R{W9} and \R{W2a}, we obtain
\bee{W22a1}
\mu S^{cab}{_{;c}} + \nu S^{[ab]c}{_{;c}}\ =\ 0,
\ee
and from \R{W22a}$_2$, \R{W10} and \R{aQ7}$_1$ follows by use of \R{W22a1}
\bee{W22a2}
\Big[\mu S^{cab}{_{;c}}+\nu S^{[ab]c}{_{;c}}\Big]_{;a}
+\lambda S^{(ab)c}{_{;c;a}}\ =\ 0\quad\longrightarrow\quad
\lambda S^{(ab)c}{_{;c;a}}\ =\ 0.
\ee
If we demand that \R{W22a1} and \R{W22a2}$_2$ are valid for arbitrary ($\lambda,
\mu,\nu$), the following conditions for the spin divergences
\bee{W22c}
S^{cab}{_{;c}}\ =\ 0,\quad S^{(ab)c}{_{;c;a}}\ =\ 0,\quad S^{[ab]c}{_{;c}}\ =\ 0
\ee
have to be satisfied for the validity of \R{W22a}. According to \R{W24}, \R{W22a}$_1$
and \R{W22a2}$_2$, we obtain the expected result
\bee{W22e}
\Theta^{ab}(2\mu-\nu)\ =\ T^{ab}
\ee
that symmetric and divergence-free energy-momentum tensors  are fix-points of the
Belinfante/Rosenfeld procedure, even if the spin is different from zero.
The special case of vanishing spin is included in \R{W22c}.

\section{Belinfante/Rosenfeld procedure and GRT\label{GRT}}

For the sequel, we now consider the case \R{W7} which causes that the curvature tensor
appears in the energy-momentum balance equation \R{W20}$_1$ which is determined
by the space-time geometry. Here, we are interested in the Riemannian space-time which
is described by Einstein's equations
\bee{b1}
R^{ab} - \frac{1}{2}g^{ab}R\ =\ \kappa\Theta^{\dagger ab}
\quad\Longrightarrow\quad
\Theta^{\dagger ab}\ =\ \Theta^{\dagger ba},\quad
\Theta^{\dagger ab}{_{;a}}\ =\ 0
\ee
(Ricci tensor $R^{ab}$, metric $g^{ab}$, curvature scalar $R$). We remind that the
following proceeding is standard\footnote{standard, but not evident}: if the energy-momentum tensor is not
symmetric\footnote{e.g. that is the case, if the stress tensor of a material is not symmetric, for liquid crystals, spin materials \C{SAX} etc.} or not divergence-free\footnote{if e.g. in General-Covariant Physics
$T^{ab}{_{;a}}\neq 0$ is valid in general},
one has to generate a symmetric and divergence-free tensor \R{W22} by use of the
Belinfante/Rosenfeld procedure. This tensor serves as RHS of \R{b1}$_1$
\bee{W24}
\Theta^{\dagger ab}\ \st{\td}{\equiv}\ \Theta^{ab}(2\mu-\nu)\ =\
T^{(ab)}-(2\mu-\nu) S^{(ab)c}{_{;c}}.
\ee
The scalars $\mu$ and $\nu$ stem from the spin balance equation \R{W20}$_2$
whose source is the anti-symmetric part of the energy-momentum tensor.
According to the setting \R{W24}$_1$, the two scalars
fix the contribution of the spin to the source of Einstein's equations, that means, the
spin's contribution to gravitation.
Using Einstein's equations and the spin balance \R{W20}$_2$, we obtain a
representation of Einstein's equations
\bee{W25}
T^{ab}\ =\ \frac{1}{\kappa}\Big( R^{ab} - \frac{1}{2}g^{ab}R\Big)
+ \Big[\mu S^{cab}+\mu S^{abc}+(\mu-\nu)S^{bac}\Big]_{;c}
\ee
which is compatible with \R{W19}$_2$ according to \R{b1}$_3$.
\vspace{.3cm}\newline
The curvature tensor in the energy-momentum balance equation \R{W20} has to be
match with the metric $g^{ab}$ in Einstein's
equations \R{b1}$_1$, that means, it has to obey the Bianchi identities as integrability
conditions \C{STE}
\bee{W26}
R^b{_{klm;j}}+R^b{_{kmj;l}}+R^b{_{kjl;m}}\ =\ 0.
\ee
Consequently, \R{W20}$_1$ becomes with $j\equiv b$
\bey\nonumber
T^{ab}{_{;a;b}} &=& \Big[R^b{_{kmb;l}}+R^b{_{kbl;m}}\Big]
\Big(\mu S^{klm}+\nu S^{[lm]k}\Big)
-R^b{_{klm}}\Big(\mu S^{klm}+\nu S^{[lm]k}\Big)_{;b}=
\\ \label{W27}
&=&\Big[-R{_{km;l}}+R{_{kl;m}}\Big]\Big(\mu S^{klm}+\nu S^{[lm]k}\Big)
-R^b{_{klm}}\Big(\mu S^{klm}+\nu S^{[lm]k}\Big)_{;b}.
\eey
\vspace{.3cm}\newline
In the next section, we investigate how the Belinfante/Rosenfeld procedure works
with regard to GCCP.

\section{GCCP and Belinfante/Rosenfeld Procedure}

Usually, a Lagrange formalism is not available for General-Covariant Continuum
Physics (GCCP). Consequently, we have to start out with the
balance equations of energy-momentum and spin which are ad-hoc
equations in the sense of a theory which is supported by a variational problem.
In a curved space-time, these balance equations are \C{B}
\byy{b5a}
T^{ab}{_{;a}}\ =\
G^b + k^b,\quad T^{ab}\ \neq\ T^{ba}, 
\qquad S^{cab}{_{;c}}\ =\ H^{ab} +m^{ab},
\\ \label{b5a1}
\mbox{with}\quad S^{cab}\ =\ -S^{cba},\quad m^{ab}\ =\ -m^{ba},
\quad H^{ab}\ =\ -H^{ba}.\hspace{1.5cm}
\vspace{.3cm}\eey
The $G^b$ and $H^{ab}$ are internal source terms {\bf--}the {\em Geo-SMEC-terms} 
(\U{Geo}metry-\U{S}pin-\U{M}o\-men\-tum-\U{E}nergy-\U{C}oupling) \C{B}{\bf--} which are caused by the choice of a special space-time geometry
and by a possible coupling between energy-momentum, spin and geometry.
For non-isolated systems, $k^b\neq 0$ denotes an external force
density, and $m^{ab}\neq 0$ is an external momentum density.
In particular, one finds such a situation in special-relativistic conti\-nuum
thermodynamics \C{B3,B1,B2}.
\vspace{.3cm}\newline
Inserting the balance equations \R{b5a}, the Mathisson-Papapetrou-like equations
\R{W20} become
\bee{W28}
G^b + k^b\ =\ -R^b_{klm}\Big(\mu S^{klm}+\nu S^{[lm]k}\Big),\qquad
T^{[ab]}\ =\ \mu\Big(H^{ab} +m^{ab}\Big)+\nu S^{[ab]c}{_{;c}}.
\vspace{.3cm}\ee
The sources of the energy-momentum and spin balance equations \R{b5a} have to
satisfy \R{W28} for performing a modified Belinfante/Rosenfeld procedure. Because
the source of the spin balance equation \R{b5a}$_3$ determines the anti-symmetric
part of the energy-momentum tensor, we demand taking \R{W28}$_2$ into account
\bee{W29}
H^{ab} +m^{ab}\ =\ 0\quad\longrightarrow\quad T^{[ab]}\ =\ 0
\quad\longrightarrow\quad\nu S^{[ab]c}{_{;c}}\ =\ 0.
\ee
Because the spin is not restricted only by fact that we apply the Belinfante/Rosenfeld
procedure with regard to GCCP, we satisfy \R{W29}$_3$ by the setting
\bee{W30}
\nu\ \st{\td}{=}\ 0
\ee
which adapts the Belinfante/Rosenfeld procedure to GCCP. Consequently,
the Mathisson-Papapetrou-like equations \R{W28} result in
\bee{W31}
T^{ab}{_{;a}}\ =\ G^b + k^b\ =\ -\mu R^b_{klm}S^{klm},\qquad
T^{[ab]}\ =\ \mu\Big(H^{ab} +m^{ab}\Big)\ =\ \mu S^{cab}{_{;c}},
\ee
These equations are similar to those generated by the general-covariant 
Belinfante/Rosenfeld procedure \R{U2}, if the balance equations of GCCP \R{b5a} are
inserted. The derivation of \R{W31} indicates that two conditions for the validity of
\R{U2} must hold: firstly the dependence of the anti-symmetric part of the
energy-momentum tensor on the source of the spin balance equation represented by
\R{W30} and secondly that the factor 1/2 in \R{U2} is introduced from knowledge
beyond the Belinfante/Rosenfeld procedure\footnote{For all special models which we
know up to now, $\mu=1/2$ is valid, so that whith \R{W30} and \R{W7}$_1$ we come
back to the usual covariant Belinfante procedure starting with \R{U1}.}.
\vspace{.3cm}\newline
From \R{W24}$_2$ and \R{b1}$_3$ follows
\bee{W32}
T^{(ab)}{_{;a}}\ =\ 2\mu S^{(ab)c}{_{;c;a}}.
\ee
Taking \R{W31}, \R{b5a}$_1$ and \R{W32} into account, we obtain a differential
equation for the spin
\byy{W33}
T^{ab}{_{;a}}\ =\ T^{[ab]}{_{;a}}+T^{(ab)}{_{;a}} &=& -\mu R^b_{klm}S^{klm}\
=\ \mu\Big(H^{ab} +m^{ab}\Big)_{;a}+2\mu S^{(ab)c}{_{;c;a}},
\\ \label{W34}
S^{(ab)c}{_{;c;a}} &=& -\frac{1}{2} R^b_{klm}S^{klm}
-\frac{1}{2}\Big(H^{ab} +m^{ab}\Big)_{;a}.
\eey
Here \R{W34} is independent of the scalar $\mu$. That is not the case
for the Belinfante/Rosenfeld generated symmetric and divergence-free tensor \R{W22}
\bee{W35}
\frac{1}{\kappa}\Big( R^{ab} - \frac{1}{2}g^{ab}R\Big)\ =\
\Theta^{ab}(2\mu)\ =\ T^{(ab)}-2\mu S^{(ab)c}{_{;c}}.
\vspace{.3cm}\ee
Also in the special case of GCCP, \R{W30}, the contribution of the spin to gravitation
is determined by knowledge from the outside of the Belinfante/Rosenfeld procedure.
The same situation holds true for the general-covariant procedure which starts out with
the definition \R{U1} resulting in \R{W16}. The Belinfante/Rosenfeld procedure
itself cannot generate the symmetric and divergence-free "mutant" \R{W35} of the
energy-momentum tensor: the scalar $\mu$ has to be known
in order to solve \R{W34} and \R{W35} together.
\vspace{.3cm}\newline
The complete formulation of GCCP contains beside the balance equations \R{b5a}
those of particle number and entropy density\footnote{For this continuum
theory of irreversible processes, see \C{B3,B3a} and the contributions in \C{B4}.}
\bee{P3}
N^k{_{;k}}\ =\ 0, \qquad S^k{_{;k}}\ =\ \sigma + \varphi
\ee
($N^k$ particle flux density, $S^k$ entropy 4-vector, $\sigma$
entropy production, $\varphi$ entropy supply). The Second Law of
Thermodynamics is taken
into account by the demand that the entropy production has to be non-negative
at each event and for arbitrary materials after having inserted the constitutive equations
into the expression of the entropy production \C{C3}
\bee{P4}
\sigma\ \geq\ 0.
\ee
Particle number and entropy density are here out of scope because the
Belinfante/Rosenfeld procedure \R{aU4} does not touch particle number and entropy.
That means, $N^k$ and $S^k$ do not influence gravitation concerning GRT.

\section{Material, GCCP and Bel/Ros Procedure}

We now presuppose that $\mu$ in \R{W35} is known from the outside, so that the
Belinfante/Rosenfeld procedure can be performed. Consequently, we consider three
tensors in GCCP$\cap$GRT
\bee{W36}
\Theta^{ab}(2\mu),\quad T^{ab},\quad S^{cab},
\ee
the source of Einstein's equations \R{W35}, the energy-momentum tensor and the
spin tensor satisfying the Mathisson-Papapetrou-like equations \R{W31}, the
energy-momentum balance \R{W31}$_1$ and the spin balance \R{W31}$_2$.
Because the Mathisson-Papapetrou-like equations are necessary for performing the
Belinfante/Rosenfeld procedure \C{A}, all energy-momentum balances and spin balances
\R{b5a} whose sources do not satisfy \R{W31} are not suitable for applying the
Belinfante/Rosenfeld procedure, and therefore also not suitable for a GRT description
because a source of Einstein's equations cannot be generated.
\vspace{.3cm}\newline
If the Mathisson-Papapetrou-like equations \R{W31} are satisfied, an energy-momentum
"mutant" $\Theta^{ab}$, \R{W35}, exists beyond the energy-momentum tensor
$T^{ab}$, and the question arises, which tensor describes the constitutive
properties of the considered system ? Evident is that in GCCP
the in general non-symmetric and not divergence-free 
energy-momentum tensor $T^{ab}$ whose (3+1)-decomposition is
\bee{P2c}
T^{ab}\ =\ \frac{1}{c^4}eu^au^b + \frac{1}{c^2}u^ap^b +
\frac{1}{c^2}q^au^b + t^{ab}
\ee
contains constitutive properties such as
\byy{P2e}
\mbox{energy density:}\quad e &=& u_ku_lT^{(kl)},
\\ \label{P2f}
\mbox{momentum flux density:}\quad p^k &=& h^k_lu_mT^{ml},
\\ \label{P2g}
\mbox{energy flux density:}\quad q^k &=& h^k_lu_mT^{lm},
\\ \label{P2h}
\mbox{stress tensor:}\quad t^{kl} &=& h^k_a h^l_bT^{ab}.
\eey
Here $u_k$ is the 4-velocity of the material and
\bee{P7}
h^i_k\ =\ \delta^i_k - \frac{1}{c^2}u^iu_k,
\ee
the projector perpendicular to the 4-velocities $u^k$ resp. $u_i$. 
\vspace{.3cm}\newline
Presupposing the
special case that the source of the energy-momentum balance equation \R{b5a}$_1$
satisfies the Mathisson-Papapetrou-like equation \R{W31}$_1$, the RHS of Einstein's
equations is \R{W35} describing the constitutive influence of the material to gravitation.
If  \R{W31}$_1$ is not valid, GCCP and GRT do not fit together. Clearly, real constitutive
properties do not change by performing a formal Belinfante/Rosenfeld procedure.
Consequently, it is evident that the energy-momentum tensor $T^{ab}$ remains a
carrier of constitutive properties also after having performed a Belinfante/Rosenfeld
procedure: a material equipped with a non-symmetric stress tensor \R{P2h} cannot be
described by a symmetric $\Theta^{ab}(2\mu)$, \R{W35}.
\vspace{.3cm}\newline
There is one special case for which constitutive properties are transferred
to the source of Einstein's equations, if $T^{ab}$ is a fix-point of the
Belinfante/Rosenfeld procedure. Starting out with \R{W35} and taking the
fix-point property into account, we obtain by use of \R{W32}
\byy{W37}
T^{(ab)}-2\mu S^{(ab)c}{_{;c}}\ =\ \Theta^{ab}(2\mu)&\doteq&
T^{ab}\ =\ T^{(ab)}+T^{[ab]},
\\ \label{W38}
\quad\longrightarrow\quad -2\mu S^{(ab)c}{_{;c}} &=&T^{[ab]}\ =\ 0
\quad\longrightarrow\quad T^{ab}{_{;a}}\ =\ 0.
\eey
The result is as expected: fix-points of the Belinfante/Rosenfeld procedure are
symmetric and divergence-free energy-momentum tensors. Consequently, the
symmetric and divergence-free "mutant" is in this case the energy-momentum tensor
itself equipped with constitutive properties.
\vspace{.3cm}\newline
Summarizing, we have three different situations:\newline
I: The Mathisson-Papapetrou-like equation \R{W31}$_1$ is not satisfied. In this case,
a GCCP-induced source of Einstein's equations cannot be generated.\newline
II: If the Mathisson-Papapetrou-like equation \R{W31}$_1$ is valid, the source of
Einstein's equations is \R{W35} which is different from the energy-momentum tensor,
the GCCP carrier of constitutive properties .\newline
III: If the energy-momentum tensor is symmetric and divergence-free,
the source of Einstein's equations is identical to this energy-momentum tensor, and
the Belinfante procedure is dispensable.
\vspace{.3cm}\newline
Indisputable, the case III correspond to the demand of GRT that all constitutive
properties are tied down to the source of Einstein's equations. This case points out
that GRT is a special constitutive theory restricted to materials of symmetric and
divergence-free energy-momentum tensors. Evident is, that in case I GCCP and GRT
do not fit together: GCCP cannot be adapted to GRT. Really interesting is the case II:
parts of the energy-momentum
tensor are considered as non-generating gravitation. We obtain this part
by the Belinfante/Rosenfeld procedure from \R{W35} and \R{W31}$_2$
\bee{W39}
T^{ab}-\Theta^{ab}(2\mu)\ =\ T^{[ab]}+2\mu S^{(ab)c}{_{;c}}\ =\ 
\mu\Big[S^{cab}+S^{abc}+S^{bac}\Big]_{;c}.
\ee
Because the formal Belinfante/Rosenfeld procedure does not change constitutive
pro\-perties, these are still described by the in general non-symmetric and
non-divergence-free energy-momentum tensor $T^{ab}$. If erroneously the constitutive
properties in case II are bound up with the symmetric and divergence-free "mutant"
$\Theta^{ab}(2\mu)$, the original material of GCCP is replaced by another one
which now fit into GRT, but which is not the original one anymore. If one reject the fact
of case II that constitutive properties and gravitational influence are described by
different tensors, we obtain the following\newline
$\blacksquare$ {\sf Statement:} Not accepting that two different tensors are necessary
for describing space-time and constitutive properties, GCCP and GRT only fit together for
special materials of symmetric and divergence-free energy-momentum tensor.
\hfill$\blacksquare$
\newline
This statement has two consequences:\newline
{\sf A:} Concerning GCCP, the Belinfante/Rosenfeld procedure is
dispensable (case III).
\newline
{\sf B:} A non-restricted general relativistic constitutive theory needs a
theory of gravitation which accepts non-symmetric and non-divergence-free
energy-momentum tensors (case I and II)\footnote{e.g. the Einstein-Cartan space-time}.

\section{Discussion}

The energy-momentum and the spin balance equations of General-Covariant Continuum
Physics (GCCP) include covariant derivatives which are determined by the space-time
geometry. Either the space-time geometry is given or the balance equations have to be
solved together with the field equations which determine the space-time geometry.
The energy-momentum tensor of GCCP is in general non-symmetric and not
divergence-free and is a carrier of constitutive properties such as energy density,
energy flux density, momentum flux density and stress tensor. 
\vspace{.3cm}\newline
If we are interested in the conditions under which GCCP and General
Relativity Theory (GRT) fit together, the first well known problem arising is, that
the source of Einstein's equations is a symmetric and divergence-free tensor.
The usual
tool for symmetrizing the energy-momentum tensor is the general-covariant
Belinfante/Rosenfeld
procedure which generates a symmetric and divergence-free tensor which can be used
as source of Einstein's equations \C{A}. 
\vspace{.3cm}\newline
The general-covariant Belinfante/Rosenfeld procedure uses an ad-hoc combination of spin divergences for achieving the symmetrization of the energy-momentum tensor.
It was found out that only such energy-momentum tensors can be symmetrized which satisfy the Mathisson-Papapetrou equations \C{A}. That means, only those materials
which satisfy the constraints implemented by the necessary
Mathisson-Papapetrou equations can be
described by GRT. Only for those materials, a symmetric and divergence-free
"mutant" of the GCCP-energy-momentum tensor can be generated by the Belinfante/Rosenfeld procedure.
\vspace{.3cm}\newline
If the above mentioned ad-hoc combination of spin divergences is replaced by a more ge\-ne\-ral family 
of spin divergences,  Mathisson-Papapetrou-like equations appear as ne\-cessary
constraints for the
Belinfante/Rosenfeld convertable energy-momentum tensors. These 
Mathisson-Papapetrou-like equations decay into two classes: one class contains the
curvature tensor, the other does not. Here, the first class is considered, because we are
interested in the relation of GCCP and GRT. The introduction of the family of spin
divergences demonstrates that there are symmetrization procedures beyond the
conventional Belinfante/Rosenfeld proceeding. In any case, conventional or modified
Bel\-infante/Rosenfeld procedure, the symmetrization needs some knowledge from the
outside of the procedure, because otherwise the result of the Belinfante/Rosenfeld
symmetrization is not unique.
\vspace{.3cm}\newline
As expected, symmetric and divergence-free energy-momentum tensors are fix-points
of the Belinfante/Rosenfeld procedures\footnote{that means, a Belinfante/Rosenfeld procedure is dispensable for such energy-momentum tensors}, even if the spin is different from zero. That is the only case for which the following problem does not appear:
constitutive properties of GCCP are tied down to the
coresponding non-symmetric and not divergence-free energy-mo\-men\-tum tensor of
GCCP, whereas the properties of the space-time geometry are generated by the
Belinfante/Rosenfeld-transformed "mutant". The formal Belinfante/Rosenfeld procedure
\R{aU4} does not transfer constitutive properties of the energy-momentum tensor to
the source of Einstein's equations\footnote{e.g. if momentum flux density and energy flux
density are different in GCCP, they become equal in GRT after having performed the
Belinfante/Rosenfeld procedure, implying a change to another material}.
\vspace{.3cm}\newline
This splitting into two tensors, the material-dependent
energy-momentum tensor and the Belinfante\-/Rosenfeld-transformed
one determining the geometry contradicts the spirit of GRT which demands that the
source of Einstein's equations should contain all constitutive properties concerning the
energy-momentum tensor. That is the case if
both these tensors are equal, that means, the energy-momentum tensor is already symmetric and divergence-free and the
symmetrization procedure is dispensable. Concerning GCCP, GRT is a theory which is
restricted to the class of materials of symmetric and divergence-free energy-momentum
tensors. Abolishing of this restriction makes an extension of the field equations beyond
GRT necessary. A possible candidate for replacing GRT with regard to GCCP may be the
Einstein-Cartan space-time geometry.
\vspace{.3cm}\newline
The question concerning the compatibility of GCCP and GRT can be answered as follows:
as expected, the compatibility exists only for materials with symmetric and divergence-free energy-momentum tensors, but only if the above mentioned spirit of GRT
is accepted.

\section*{Epilogue}

In the special-relativistic Belinfante-Rosenfeld symmetrization  
procedure of a divergence-less non-symmetric energy-momentum tensor $T^{ab}$,
the following ansatz
\bey\nonumber
^\dagger\Theta^{ab}\ :=\ T^{ab}
-\frac{1}{2}\Big[S^{cab}+S^{abc}+S^{bac}\Big]_{,c}\hspace{4cm}(*)
\hspace{-4.5cm}
\eey
 is assumed. Here the antisymmetry of $S^{cab}$ in the last two indices  
 $a$ and $b$ guaranties that 
\bey\nonumber
\Sigma^{cab}\ :=\ S^{cab} + S^{(ab)c}+ S^{[ab]c}
\eey
 is anti-symmetric in the first two  
 indices $c$ and $a$ such that, due to the commutativity of partial  
derivatives, $T^{ab}$ can be symmetrized by (*).
\vspace{.3cm}\newline
 In the general-covariant derivation of the Mathisson-Papapetrou  
 equations given in \C{A},  the starting point is  the symmetrization of  
 a non-symmetric energy-momentum tensor, whose divergence does not  
 vanish.To this end, the general-covariantly generalized ansatz \R{U1} is  
 assumed what, to some extend, is justified by the special- 
 relativistic Belinfante-Rosenfeld procedure. But, the strategety of  
 the present paper is the following: We do not refer to our knowledge  
 of special relativity, but ask\newline 
{\bf(i)} whether there are other general-covariant algebraic combinations of the
tensor $S^{cab}$ --now specified as the spin tensor of continuum physics
according to \R{b5a}$_3$-- which symmetrize $T^{ab}$  and\newline 
{\bf(ii)} how other combinations of the spin tensor modify the Mathisson-Papapetrou equations.\newline
 The result is that for the class of ansatzes \R{aQ7}, up to a factor $\mu$, \R{U1}  
 is reproduced, if additional assumptions are made. This corroborates our knowledge
from special relativity that only the symmetry properties of $\Sigma^{cab}$ are  
relevant for the symmetrization procedure. While these properties are  
not changed by $\mu$, the Mathisson-Papapetrou-like equations \R{W31}$_2$  
contain this arbitrary factor which has to be determined otherwise.
\vspace{.8cm}\newline
{\bf Acknowledgement} My warm thanks to Prof.Dr. H.-H. v. Borzeszkowski, Institut
f\"ur Theoretische Physik, TU Berlin, for numerous helpful and sometimes controversial,
but always friendly discussions on the meaning of the Mathisson-Papapetrou equations
in the context with the general-covariant Belinfante/Rosenfeld procedure and the
general-covariantly formulated Continuum Physics. Not to get lost of different points of
view, we decided to add the epilogue.

\end{document}